\documentclass[pra,aps,preprint]{revtex4-1}
\usepackage{graphicx}
\usepackage{amsmath,amssymb}
\usepackage{dsfont}

\newcommand{\sts}{\vert_{t\rightarrow\infty}}
\newcommand{\Hq}{\mathcal{H}}

\newcommand{\hsl}{\mathcal{H}_{\text{\tiny SL}}}

\newcommand{\nn}{\nonumber}

\newcommand{\wo}{\omega_{\circ}}

\newcommand{\rl}{\rho_{L}^{eq}}

\newcommand{\ic}{\vert_{t=0}}

\begin{document}
\title{Dissipative phase transition in a spatially-correlated bosonic bath}
\author{Saptarshi Saha}
\author{Rangeet Bhattacharyya}
\email{rangeet@iiserkol.ac.in}
\affiliation{Department of Physical Sciences, Indian Institute of Science Education and Research Kolkata,\\
Mohanpur -- 741246, WB, India}

\begin{abstract}
The presence of symmetries in a closed many-body quantum system results in integrability. For such
integrable systems, complete thermalization does not occur. As a result, the system remains non-ergodic. On
the other hand, a set of non-interacting atoms connected to a regular bosonic bath thermalizes. Here, we
show that such atoms in a spatially-correlated thermal bath can show both the behavior depending on the
temperature. At zero temperature, the bath has a large correlation length, and hence it acts as a common
environment. In this condition, a set of weak symmetries exist, which prevent thermalization. The system
undergoes a symmetry-broken dissipative phase transition of the first order as the temperature rises above
zero.
\end{abstract}

\maketitle

\section{Introduction}

The irreversible journey of closed or open quantum systems towards thermal equilibrium,
i.e., the thermalization problem, has been the subject of intense research for many
decades. For a set of classical particles, Boltzmann suggested an explanation based on the collisions
\cite{Kardar2007}. Later, in 1929, von Neumann proposed the quantum ergodic theorem, which is the first step
towards the quantum thermalization problem \cite{von_neumann_proof_2010}. About three decades ago, Srednicki
proposed the famous eigenstate-thermalization hypothesis (ETH) for a closed many-particle system. According
to ETH, a specific state of a closed interacting many-particle quantum system would thermalize, provided
that state's energy eigenstate representation obeys Berry's conjecture \cite{srednicki_chaos_1994}. For such
a system, the ensemble average of an observable reaches the thermal expectation value after a long time. One
can partition such a system into a chosen subsystem, and an effective heat bath from the remainder of the
system \cite{nandkishore_many-body_2015}. Consequently, the von Neumann entropy of the subsystems is
extensive, and it follows the volume law \cite{abanin_colloquium_2019}.

Closed many-body systems may not thermalize if they are in a localized state. Anderson showed the
non-thermal nature of the disordered systems, known as Anderson localization \cite{anderson_absence_1958}.
There exist several integrals of motion from the symmetry of the system, which break the principle of equal
apriori probabilities.  Therefore, integrability remains one of the measures of the non-thermal phases. Many-body
localization (MBL), quantum scars are the examples where the ETH fails due to the presence of symmetries
\cite{abanin_colloquium_2019, Pal_2010, Oganesyan_2007, Turner_2018}. MBL eigenstates are known to follow
the area-law entanglement in place of the volume law.  The subsystems become entangled in the localized
phase \cite{abanin_colloquium_2019}.

On the other hand, the system is coupled to a thermal bath with a specific
temperature for the open quantum system. The system-bath interaction remains the source of thermalization, and the quantum master
equation describes the reduced dynamics of the system \cite{Breuer_2002, kossakowski_quantum_1972}. The
system inherits the temperature from the bath. For example, the Bloch equation for spin-$1/2$ particles
successfully describes the dynamical evolution toward equilibrium configuration \cite{Bloch_1946}.

In dissipative quantum systems, transitions can happen between thermal to non-thermal phases.
The steady-state solution of the master equation is, in general, a function of the parameters of
the system and the bath. At a critical value of the parameters, the steady-state configuration may undergo
a sudden change. The more detailed analysis involves the study of the eigenvalues of the Lindbladian. The
thermal steady-state corresponds to the zero eigenvalue of the Lindbladian. The eigenvalue with the smallest non-zero
absolute value provides the asymptotic decay rate (ADR), which determines the rate of approach to the
steady-state. If a continuous change of a system parameter results in the vanishing of the ADR and an
emergence of a dark state, then such a change is termed a dissipative phase transition (DPT)
\cite{kessler_dissipative_2012, albert_symmetries_2014, horstmann_noise-driven_2013, buca_note_2012,
manzano_symmetry_2014, minganti_spectral_2018, lieu_symmetry_2020}.  The vanishing of ADR ensures that the
dark state does not evolve under system-bath coupling Hamiltonian \cite{buca_note_2012}.  As such, such
states are often described as a decoherence-free sub-space in quantum optics and quantum computation
\cite{fleischhauer_quantum_2002, mohapatra_giant_2008}.  DPT is closely related to the symmetry-breaking
transitions \cite{albert_symmetries_2014, lieu_symmetry_2020, manzano_symmetry_2014}. A thermal phase is a
symmetry-broken phase due to its lack of integrability. For such systems, the final steady-state is unique,
and all the memory of initial states are lost \cite{nandkishore_many-body_2015}. On the other hand, the
non-thermal phases are protected by several symmetry operators. Hence the final steady-state in that case
has the initial value dependence, and there exist different integrals of motions along with the total energy
\cite{kessler_dissipative_2012}. 

In this work, we show that a spatially-correlated bosonic bath can connect the two extreme cases described
above.  We consider a set of non-interacting quantum systems weakly coupled to a spatially-correlated
bosonic bath.  The bath correlation function generally decays over finite length for a spatially-correlated
bath \cite{jeske_derivation_2013, mccutcheon_long-lived_2009}. At zero temperature, the bath acts as a
common environment. Hence, a cooperative effect between a pair of systems is observed in the form of an
entanglement \cite{braun_creation_2002, benatti_environment_2003, carmichael_analytical_1980}. The
bath-induced entanglement helps the system to escape the thermal steady state. We also identify the weak
symmetry operators, which are preserved during the dynamics. As the temperature increases, the bath's
correlation length becomes shorter, and the spin-pairs are disentangled. As the entanglement vanishes, the
system reaches a thermal Gibb's state. We find out the weak symmetry broken phase transition at the critical
point of the temperature. We find that the phase transition is similar to the thermal-to-localized phase
transition observed in closed many-body systems. There is the possibility of getting a non-thermal phase at
non-zero temperature if we make the bath correlation length much longer.  Reservoir engineering, which
involves adjusting the parameters of a bath, is an efficient technique for creating such a non-thermal
atomic state in optical cavities and ion trap experiments \cite{Poyatos1996, Plenio2002, Bose1999}. With the
increase in the number of atoms, the number of integrable quantities increases, and the von Neumann entropy
is no longer extensive.      

\section{Non-interacting systems in a spatially-correlated bath}

We consider two non-interacting spins coupled to a bosonic bath. We shall generalize the
result to multiple systems later in the manuscript.
The total Hamiltonian for these systems and the bath is given by,
\begin{eqnarray}
\Hq =\sum\limits_{m=1}^2\Hq_{ms}^0 +\Hq_{L}^0+\hsl
\label{eq:1}
\end{eqnarray}
where, $\Hq_{1s}^0,\Hq_{2s}^0$ are the Zeeman Hamiltonians of the spins. $\Hq_{(1,2)s} =
\omega^{(1,2)}_0 \sigma_3^{(1,2)}/2$, where, $\wo$ is the Larmor frequency and $\sigma_i$s are the
Pauli spin matrices.  $\Hq_{L}^0$ is the free Hamiltonian of the bosonic bath, given by
$\Hq_{L}^0=\sum\limits_k \omega_k a^{\dagger}_ka_k$. The spins are weakly-coupled to the bath. We
use the coupling Hamiltonian $\hsl$ adapted from a spatially-correlated spin-boson model
\cite{mccutcheon_long-lived_2009}. As such, we have $\hsl = \sum\limits_{m=1}^2
(\sigma^m_+\mathcal{L}_m+\sigma^m_-\mathcal{L}^{\dagger}_m )$, where $\mathcal{L}_m$ is the
corresponding bath modes. The bath modes are defined as, $\mathcal{L}_m(r)= \sum\limits_k g_k^m
a_k$. Here, $g_k^m$ is the coupling amplitude of the $m$th spin and the bosonic bath and is given by
$g_k^m= g_ke^{ikr_m}$. Hence, $g_k^1 \neq g_k^2$ but $\vert g_k^1 \vert = \vert g_k^2 \vert$. Two
spins has a spatial separation, $ d = \vert r_1-r_2 \vert$.  In the interaction representation of
the free Hamiltonian, the dynamical equations of the reduced density matrix of the spins is given by
the coarse-grained quantum master equation (QME) \cite{Breuer_2002}. 
%The bath is in
%thermal equilibrium, so the equilibrium density matrix is given by, $\rl= e^{-\beta
%\Hq_{L}^0}/\mathcal{Z}_L$. Where $\mathcal{Z}_L$ is the partition function of the bath and $\beta$ is the
%inverse temperature of the bath, so $T = 1/\beta$.  The correlation between the system and bath at the
%beginning of the course-graining interval is neglected (Born-Markov approximation) \cite{atom-photon}.  
The bath is assumed to be in thermal equilibrium,
so the equilibrium density matrices are given by, $\rl= e^{-\beta \Hq_{L}^0}/\mathcal{Z}_L$, where,
$\mathcal{Z}_L$ is the partition function of the bath, and $\beta$ is the inverse temperature of the bath,
so $T = 1/\beta$.  The initial correlation between the system and bath are neglected (Born-Markov
approximation) \cite{atom-photon}.  
Also, it follows from the definition of $\hsl$ that ${\rm Tr}_{\text{\tiny
L}}\{ \hsl\rl\} = 0$, which ensures that the system-local environment coupling only contributes in the
second-order.

After Jeske and others, we consider a one-dimensional chain of spatially located, coupled harmonic
oscillators, the ``tight-binding chain" as was originally named by by the authors and the spatial
correlations decay over a characteristic correlation length $\xi = 2\omega \beta g_b$, resulting from the
excitations hopping along the chain and thermal noise. Here, $\omega$ is the bath lattice spacing,
$g_b$ is the coupling between the neighboring bath oscillators, and $\beta$ is the inverse temperature
\cite{jeske2013}.  We assume both the temporal and spatial bath correlations are stationary in time and
space \cite{jeske2013}.  Bath spectral density functions corresponding to $\sigma^{n}_{-}\sigma^{m}_{+}$ and
$\sigma^{n}_{+}\sigma^{m}_{-}$ terms (The second-order terms of $\hsl$ ), for $n \neq m, \{n,m\}\in
\{1,2\},$ in the dissipator are, respectively, given as,
\begin{eqnarray}
\Gamma_{mn}(\omega_{\circ},r) &=&\sum\limits_{k} g_k^m {g_k^n}^{\star} \int\limits_0^{\infty} dt  e^{-i(\omega_{\circ}-
\omega_k)t}\langle a^{\dagger}_k a_k \rangle e^{-i k.r}  \nn\\
\Delta_{mn}(\omega_{\circ},r) &=& \sum\limits_{k} g_k^n {g_k^m}^{\star} \int\limits_0^{\infty} dt  e^{i(\omega_{\circ}- 
\omega_k)t}\langle a_k a^{\dagger}_k \rangle 
e^{-i k.r}  
\end{eqnarray}

In deriving the above, one encounters a double sum over $k$ and $k^{\prime}$ in the spectral density
functions and $k \neq k^{\prime}$ cases vanish owing to differences in $\omega_k \neq \omega_{k^{\prime}}$ a
situation akin to rotating wave approximation. For a bosonic bath, we have $\langle a_k a^{\dagger}_k \rangle= (1+N(\omega_k))$ and
$\langle a^{\dagger}_k a_k  \rangle=  N(\omega_k)$, where, $N(\omega_k)=1/(\exp(\beta \omega_k)-1)$.

To
simplify the representation of the dynamical equation in the following, we also define $A_{mn}(\wo,r)=
\Gamma_{mn}(\wo, r ) + \Gamma_{mn}^* (\wo, r )$, $B_{mn}(\wo,r)= \Delta_{mn}(\wo, r) + \Delta_{mn}^* 
(\wo, r )$ and
$J_{mn}(r)= \frac{1}{2i} (\Gamma_{mn}(\wo, r ) - \Gamma_{mn}^* (\wo, r))$, $K_{mn}(r)= \frac{1}{2i} 
(\Delta_{mn}(\wo,r ) - \Delta_{mn}^* (\wo, r ))$.  

For $n=m$, one obtains terms similar as the above, except the spectral densities do not have
the $r-$dependent $e^{-i k.r}$ terms.  In this limit, the bath spectral density functions contain a sum of
Dirac deltas, $\sum\limits_k \vert g_k\vert^2\delta(\wo-\omega_k) $. The spectral density terms, in
this case, characterize a local environment and not a common environment.

We expect that the cross-terms from the common environment should vanish when the two qubits are far
apart ($r \ll \xi$), i.e., when the temperature is high ($\beta \to 0$) corresponding to short
spatial correlation length. On the other hand, when the two
qubits are close ($r \gg \xi$), or the temperature goes to zero ($\beta \to \infty$), the common
environment effect is the highest \cite{mccutcheon_long-lived_2009}. In line with the previous works, it
had been assumed that a $r$-dependent function (say, $\alpha(r)$ with $r = \vert \vec{r}_1 -
\vec{r}_2\vert$) could be chosen to scale the spectral density functions.  This function must satisfy
$\alpha(\infty) = 0$, which ensures that the cross-terms from the common environment vanish when the
two qubits are very far apart and $\beta \to \infty$ , and $\alpha(0) = 1$, which ensures the maximum extent of
the common environment effect. In earlier work, Jeske and others have proven that any dissipator with
exponential or Gaussian spatial correlation function could be mapped to Lindblad form, ensuring CPTP. This
allows the choice of an exponential form of $\alpha(r) = \exp(-r/\xi)$, where, $\xi$ is a
measure of the spatial correlation length of the bath \cite{jeske2013}. The correlation length is
proportional to $\beta$ \cite{mccutcheon_long-lived_2009}.

We also use this result and note that $C(r) = e^{-r/\xi}C(0) = \alpha(r)C(0), C\in\{A,B,J,K\}$ where, $C(0)$ is
functions of the Zeeman frequency $\wo$, but is not a function of $r$ \cite{jeske2013}. Here $\xi$ is
the previously-defined bath correlation length. $\alpha$ is a measure of how the local bath between the neighboring
atoms has overlap or spatially correlated. For $r/\xi \to 0$, we obtain $\alpha \to 1$, and the qubits
completely share the common bath. On the contrary, for $r/\xi \to \infty$ corresponds to $\alpha \to 0$. In
this case, each spin relaxes with its own rates, and the Lindbladian does not have any cross term between
the spins.  As such, the atoms are separated, and the common environment effect is absent. Between these
two limits, the spins are in a partially overlapping bosonic bath. 

%We note that the bath correlation length
%$\xi$ is expected to be a function of the bath temperature. Intuitively, it is expected that $\xi$ would be shorter
%at higher temperatures and vice versa. As such, although we do not impose a specific functional dependence,
%yet we can study at least one extreme limit of the correlation length $\xi\rightarrow\infty$ (or, $\alpha
%\rightarrow 1$) to correspond to zero temperature. 

The form of the GKLS eq. is given by,
\begin{eqnarray}
\label{qme}
\frac{d \rho_s}{dt} &=& -i[\mathcal{H}_{lamb},\rho_s]+ \mathcal{D}\rho_s
\end{eqnarray}
Here, $\mathcal{H}_{lamb}$ is the second order shift term due to system-local environment coupling
Hamiltonian, and is given by,
\begin{eqnarray}
\mathcal{H}_{lamb}&=& \sum\limits_{i,j=1}^2\left( J_{ij} \mathcal{O}_{+}^i\mathcal{O}^j_- - K_{ij} \mathcal{O}_-^i\mathcal{O}_+^j \right)
\end{eqnarray}

Here, $J_{ij} = J(0)$ for $i=j$ and $J_{ij} = J(r)$ for $i=j$  and the notation is same for $K$ and
$\mathcal{O}_+^1=\sigma_+\otimes \mathds{1}$, $\mathcal{O}_+^2=\mathds{1}\otimes\sigma_+ $.  The Lamb shift
from the common environment have been first reported by McCutcheon and others, and this result exactly matches
with the earlier report \cite{mccutcheon_long-lived_2009}.  The form of the dissipator, $\mathcal{D}\rho_s$
is obtained as,

\begin{eqnarray}\label{DHsl}
\mathcal{D}\rho_s &=& \sum\limits_{i,j=1}^2 \Big[ B_{ij}\big(2 \mathcal{O}_-^i \rho_s \mathcal{O}_+^j -\{ \mathcal{O}_+^j \mathcal{O}_-^i , \rho_s\} \big) \nn\\ 
&& + A_{ij}\big(2 \mathcal{O}_+^i \rho_s \mathcal{O}_-^j -\{ \mathcal{O}_-^j \mathcal{O}_+^i , \rho_s\} \big)\Big]
\end{eqnarray}

The notation of $A_{ij}, B_{ij}$ is the same as $J_{ij},K_{ij}$.  The natural choice for analyzing the master
equation described by Eq. (\ref{qme}), is to move to a Liouville space description. 
Liouville space presentation of the QME is, $\left[\frac{d \hat{\rho_s}}{dt} = \mathcal{\hat{L}}\hat{\rho_s}
\right]$ where, $\mathcal{\hat{L}}$ is the Liouvillian superoperator. 
The resulting
Liouvillian matrix is a $n^2 \times n^2$ matrix, and the density matrix is a $n^2 \times 1$ column
matrices, where $n$ is the length of the Hilbert space and grows exponentially with the number of systems.
The size of the Liouvillian is not in a convenient form for
algebraic manipulation. We note that the Liouvillian is completely symmetric with respect to exchanging
the labels of spins 1 and 2.  Instead of using 15 independent elements of the reduced density matrix $\rho$, we 
use the observables'
expectation values constructed from Pauli matrices of the qubits. The equations (\ref{observables}) describe
the construction of the observables and their expectation values. We note that the positive (negative) signs
generate a set of symmetric (asymmetric) observables with respect to the exchange of qubit indices. It is
clear that we have nine symmetric and six asymmetric observables, which are given by,

\begin{eqnarray}
M_{\alpha}^{(\pm)} &=& \frac{1}{2} {\rm Tr}_s  [ (\sigma_{\alpha}\otimes \mathbb{I} \pm \mathbb{I} \otimes \sigma_{\alpha})\rho_s ] \nn\\
M_{\alpha\beta}^{(\pm)}&=& \frac{1}{4} {\rm Tr}_s [(\sigma_{\alpha} \otimes \sigma_{\beta} \pm \sigma_{\beta} \otimes \sigma_{\alpha})
\rho_s ],\quad \forall\ \alpha\neq\beta \nn \\
M_{\alpha\alpha} &=& \frac{1}{4} {\rm Tr}_s [(\sigma_{\alpha}\otimes \sigma_{\alpha})\rho_s] 
\label{observables}
\end{eqnarray}  

where, $\alpha,\beta \in \{x,y,z\}$. We construct the equations for the above nine observables by using the
dissipators used in the previous section. We note that for a Liouvillian that remains invariant with respect
to an exchange of spin qubit indices, the asymmetric observables remain zero throughout the dynamics and are
ignored in the rest of the manuscript.  The symmetric observables are denoted without the superscript $(+)$
in the manuscript's remaining part. In terms of observables, we can write the Eq.-\ref{qme} as inhomogeneous
first-order coupled linear differential equations. These equations are  Bloch-type equations for a two-spin
system \citep{Bloch_1946}. 

\begin{widetext} 
The dynamical equations for the set $\{M_z, M_{zz},
M_c \}$ are found to be,
\begin{equation}\label{3eq}
\left[\begin{array}{c} \dot{M}_z \\ \dot{M}_{zz}\\ \dot{M}_{c} \end{array} \right] = \begin{bmatrix}
-2R_1  & 0 & 4M_{\circ}\alpha R_1 \\   M_{\circ}R_1& -4R_1  & 2 \alpha R_1  \\ 
-M_{\circ} \alpha R_1 & 4 \alpha R_1  & -2R_1  \end{bmatrix}  \left[\begin{array}{c} M_z  \\
M_{zz}\\ M_c \end{array}\right] + \left[\begin{array}{c} 2M_{\circ} R_1 \\  0\\ 0 \end{array}\right]  
\end{equation} 
where, $R_1 = A(0)+B(0)$, $M_0=\frac{B(0)-A(0)}{B(0)+A(0)}$, $\alpha= e^{-r/\xi}$ from the definitions. 
Here $M_c = M_{xx}+M_{yy}$ 
\end{widetext}

\section{Dissipative phase transition}

We begin with an analysis of the eigenvalues of the Lindbladian mentioned in the previous section.
The final steady-state density matrix is an eigenvector of the Lindbladian matrix with zero
eigenvalue. An open quantum system approaches the equilibrium state with a characteristic timescale
given by the eigenvalue with a negative real part and minimum absolute value. It is also called
asymptotic decay rate (ADR) \cite{albert_symmetries_2014, horstmann_noise-driven_2013}. 
Horstman and others have shown that the eigensystem of the Lindbladian contains valuable details on the
system dynamics \cite{albert_symmetries_2014, horstmann_noise-driven_2013}. Figure-1 shows the
eigenvalues of the Lindbladian, constructed as described before, in the limits of $\xi \neq 0$
(figure (a)) and $\xi = 0$ (figure (b)). In (a), the ADR arises from the lowest absolute decay
rate shown using a blue marker (on the negative real axis). As $\xi$ increases (a signature of
lower temperature), this eigenvalue approaches zero. At $\xi \rightarrow \infty$, i.e., at the zero
temperature, this eigenvalue vanishes (shown in the figure (b)). 

\begin{figure}[htb]
\raisebox{3cm}{\large\textbf{(a)}}\hspace*{-1mm}
\includegraphics[width=0.45\linewidth]{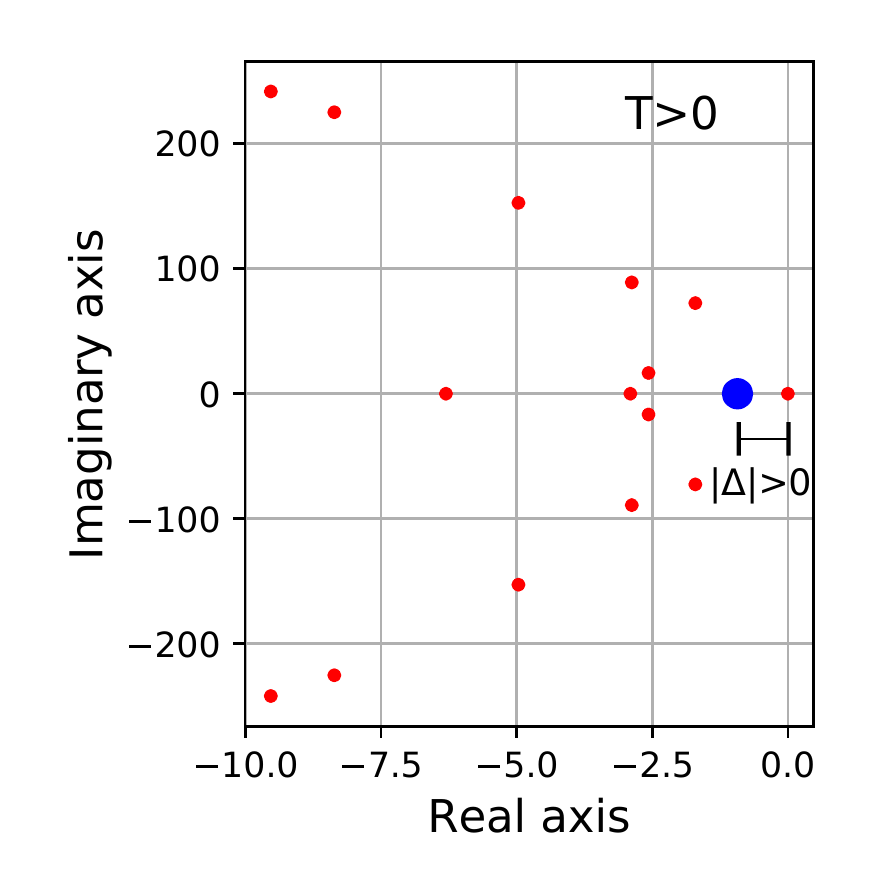} 
\raisebox{3cm}{\large\textbf{(b)}}\hspace*{-1mm}
\includegraphics[width=0.45\linewidth]{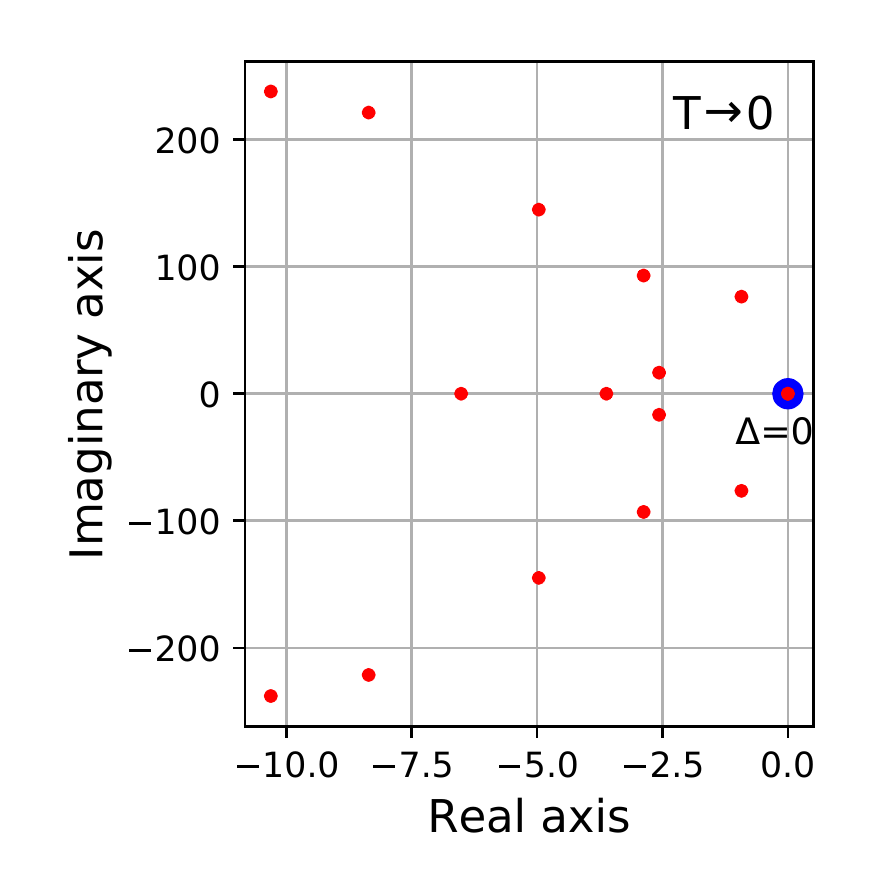}
\caption{Numerical plots for the distribution of eigenvalues in the real and imaginary axis for two different
temperatures. (a) For non-zero temperature, the only one eigenvalue is zero, and $\Delta \neq 0$. The lowest
absolute eigenvalue -- on the negative real line -- is denoted by a large blue marker, whereas other
eigenvalues are depicted by small red markers (color online).
(b) For $T = 0$, we have $\Delta=0$, i.e., the ADR vanishes and we have a frozen non-thermal state.}
\end{figure}

The quantum phase transition (QPT) occurs when as there is a level crossing between the ground state and the
first excited state for some critical value of a suitable parameter \cite{kessler_dissipative_2012,
Sachdev_1998}.  On the other hand, the DPT is associated with an explicit symmetry breaking, as the
degeneracy of the steady-state vanishes by crossing the critical limit \cite{buca_note_2012}.  The DPT
arises when the spectral gap $\Delta$ vanishes \cite{kessler_dissipative_2012}.  The Liouvillian is then
degenerate. For an ensemble of two spin-system, if there exists a parameter $\alpha$ such that, beyond a
critical value of $\alpha$ (say, $\alpha_c$), the transition between two different phases takes place.
Hence, for $\alpha<\alpha_c$, one eigenvalue of $\mathcal{\hat{L}}$ is zero, and for $\alpha \geq\alpha_c$,
more than one eigenvalues are zero.  When $T>0$, the Liouvillian has only one zero eigenvalue state. As the
trace is preserved in the whole dynamics, the steady-state dynamics is confined to the observables
$\{M_z,M_{zz}\}$ in the Eq.  (\ref{3eq}). The steady-state solution is given by,

\begin{eqnarray}\label{3sol-neq1}
M_z\sts &=& M_{\circ} \nonumber \\
M_{zz}\sts &=& M_{\circ}^2/4 
\end{eqnarray}

The steady state density matrices is thermal, $\rho^{ss}_s = e^{-\beta
(\mathcal{H}^0_{1s}+\mathcal{H}^0_{1s})}/\mathcal{Z}_s$, where $\mathcal{Z}_s$ is the partition function of
the system. Consequently, the equilibrium magnetization is $M_{\circ}= \tanh(\beta \omega_{\circ}/2)$.  On
the other hand, for $T = 0$, there exist a weak symmetry generator $\hat{D}, [S=\sigma_{x}\otimes
\sigma_{x}+\sigma_{y}\otimes \sigma_{y}+\sigma_{z}\otimes \sigma_{z}$] and under the unitary transformation
[$\hat{U}(t)=\exp(-i\hat{D}t)$]\, $\mathcal{L}$ is conserved. Following Noether's theorem, for every
symmetric operation, there exist a conserved quantity \cite{albert_symmetries_2014}.  Moreover, a symmetry
operator $\mathcal{P}$ is strong if it commutes with each of the Lindbladian operators (such as,
$\mathcal{O}$ in the Eq. (\ref{DHsl})), and is weak if it commutes only with the entire Liouvillian
\cite{lieu_symmetry_2020}.  From the Eq. (\ref{3eq}), we obtain,
\begin{eqnarray}
\frac{d}{dt}\left( M_c+M_{zz} \right) = 0.
\end{eqnarray}   

Therefore, the final solution has an initial value dependence. $\mathcal{\hat{L}}$ can be written as a
block-diagonal form of the eigenbasis of the superoperator of $\hat{D}$.  The steady-state dynamics is
confined in the observables-$\{M_z,M_{zz},M_c\}$. 
The steady state solution is given by,

\begin{eqnarray}\label{3sol-eq1}
M_z\sts &=& M_{\circ} \left[   \frac{ (4 F + 3)}{      
\left(M_{\circ}^2+3\right)} \right] \nonumber \\
M_c\sts &=& \left[\frac{4 F -  M_{\circ}^2  }{2     
\left(M_{\circ}^2+3\right)}\right] \nonumber \\
M_{zz}\sts &=& F - M_c 
\end{eqnarray}

where, $F =(M_{zz}+M_{c})\ic$. This result is in agreement with earlier work \cite{benatti2006b}. The
physical meaning of $M_{c}$ can be emphasize in a simpler manner. Using the expression for the steady state
concurrence, $C(\rho_s^{ss})$, the condition for the persistent entanglement is given by, $4 \vert M_c
\vert>\sqrt{(1+4M_{zz})^2-4M_z^2}$  \cite{hill_entanglement_1997}. Therefore, if $M_c \neq 0$, it signifies
there is a possibility of bath-induced persistent entanglement in this phase. We note that the bath induced
entanglement (for a common bath) is a well-known concept since the seminal work of Plenio and others
\cite{plenio_cavity-loss-induced_1999, an_entanglement_2007, benatti_environment_2008,
zhang_entanglement_2007, choi_quantum_2007}.

\begin{figure*}[tb]
\raisebox{3cm}{\large\textbf{(a)}}\hspace*{-1mm}
\includegraphics[width=0.28\linewidth]{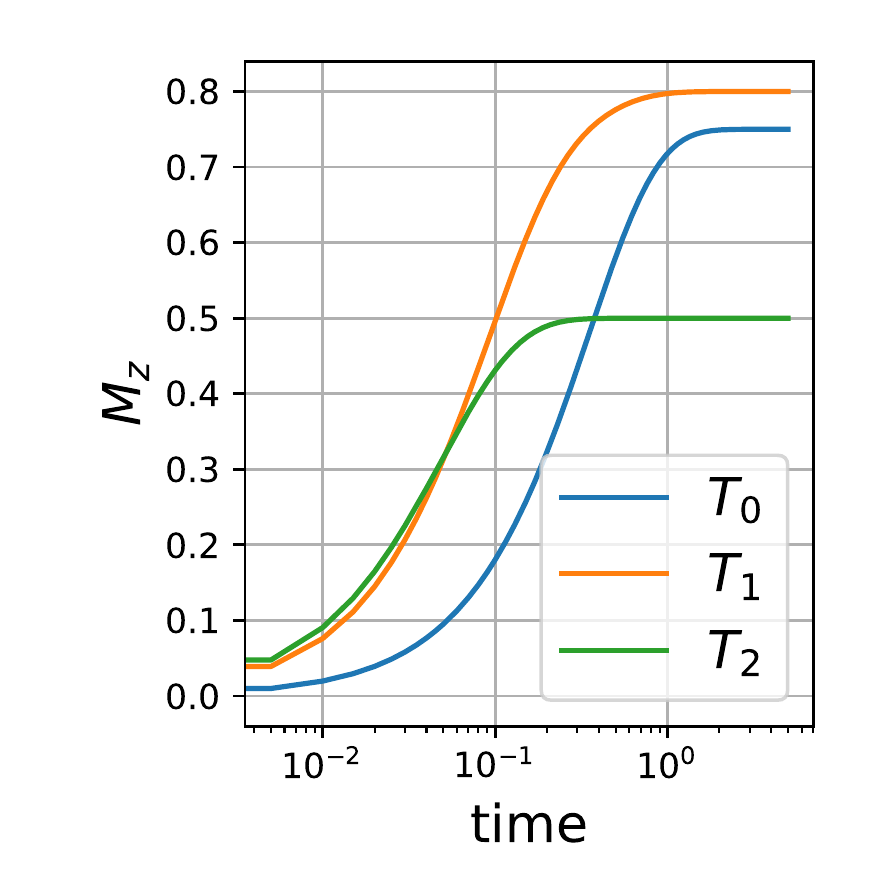} 
\raisebox{3cm}{\large\textbf{(b)}}\hspace*{-1mm}
\includegraphics[width=0.28\linewidth]{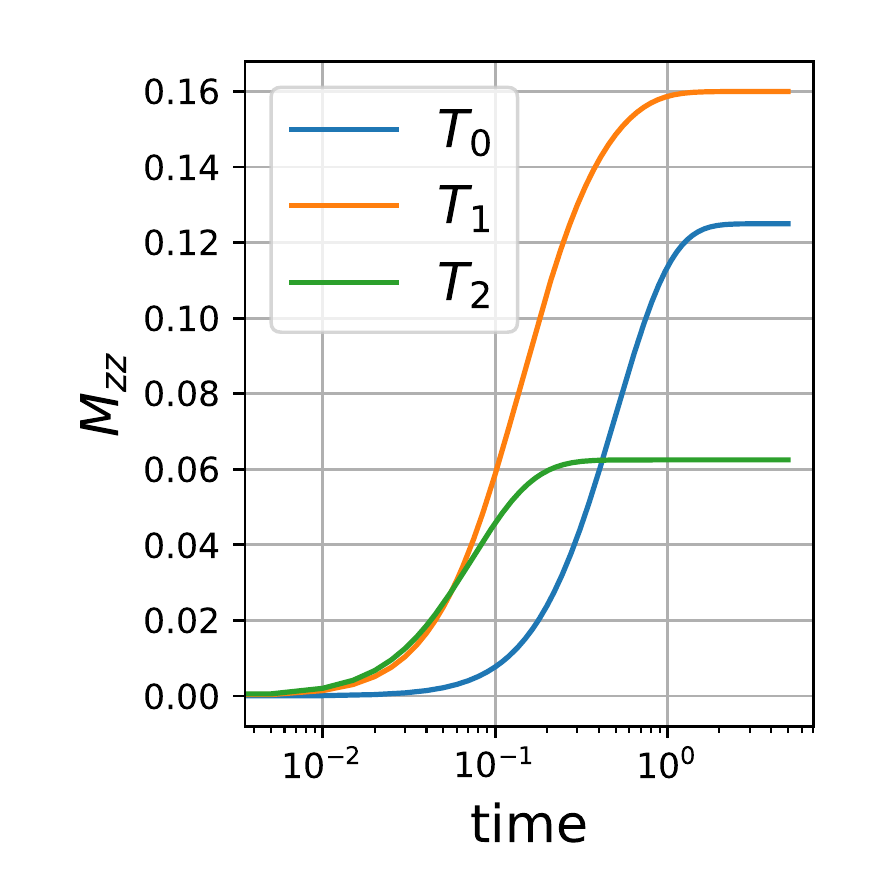}
\raisebox{3cm}{\large\textbf{(c)}}\hspace*{-1mm}
\includegraphics[width=0.28\linewidth]{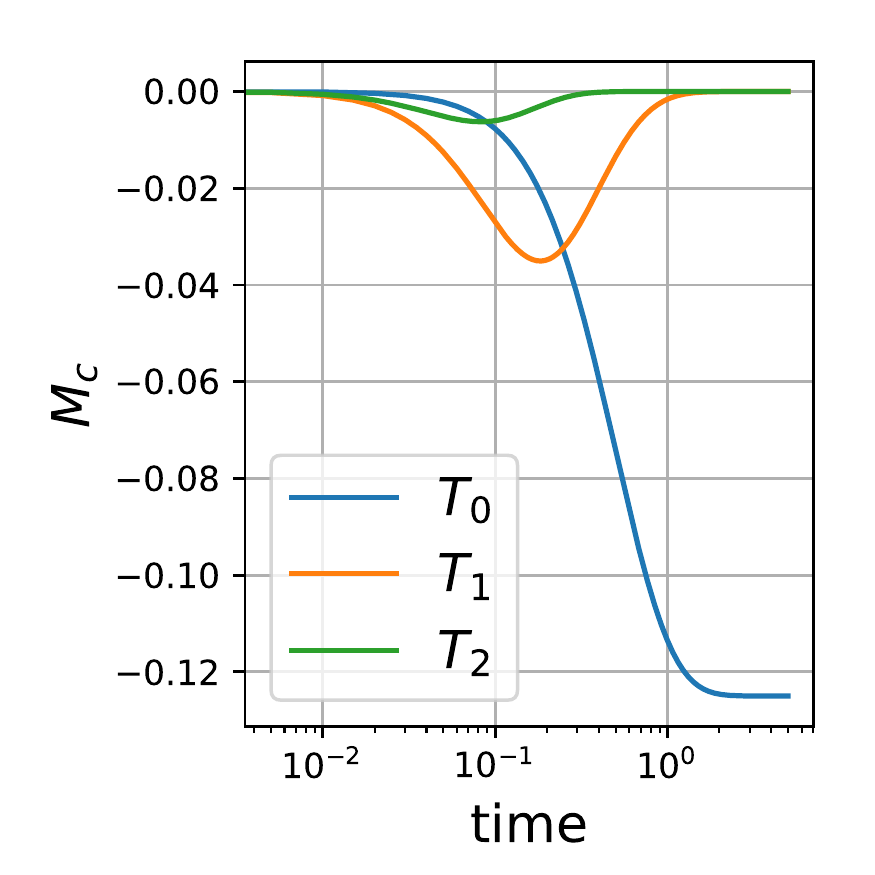}\\
\raisebox{3cm}{\large\textbf{(d)}}\hspace*{-1mm}
\includegraphics[width=0.28\linewidth]{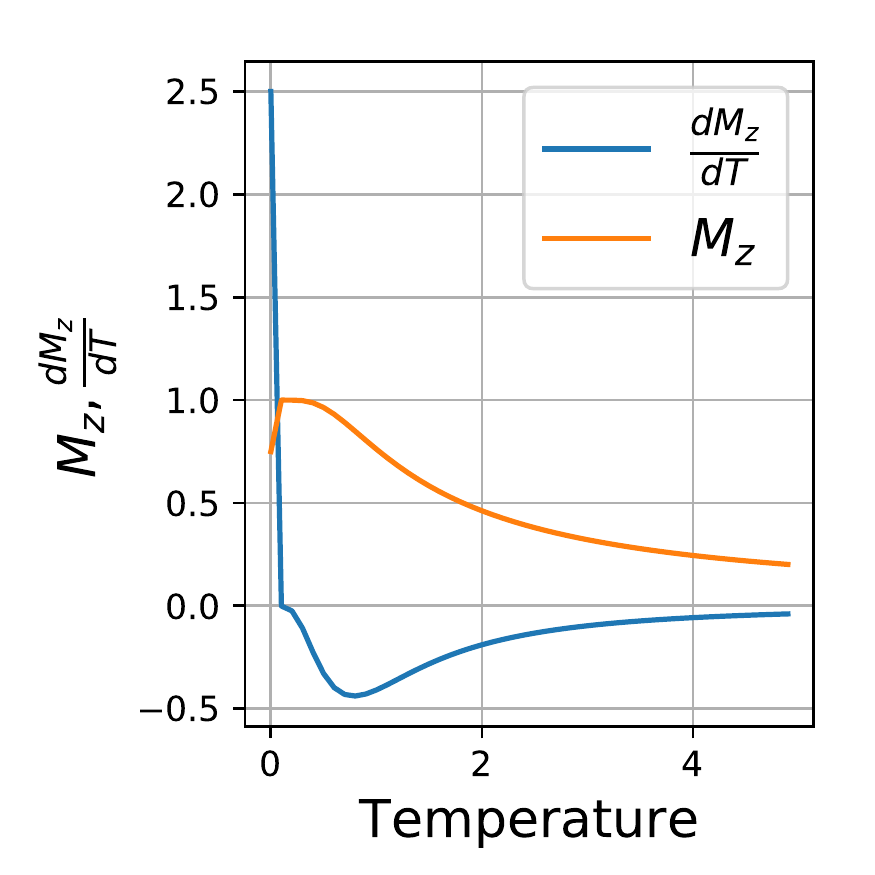} 
\raisebox{3cm}{\large\textbf{(e)}}\hspace*{-1mm}
\includegraphics[width=0.28\linewidth]{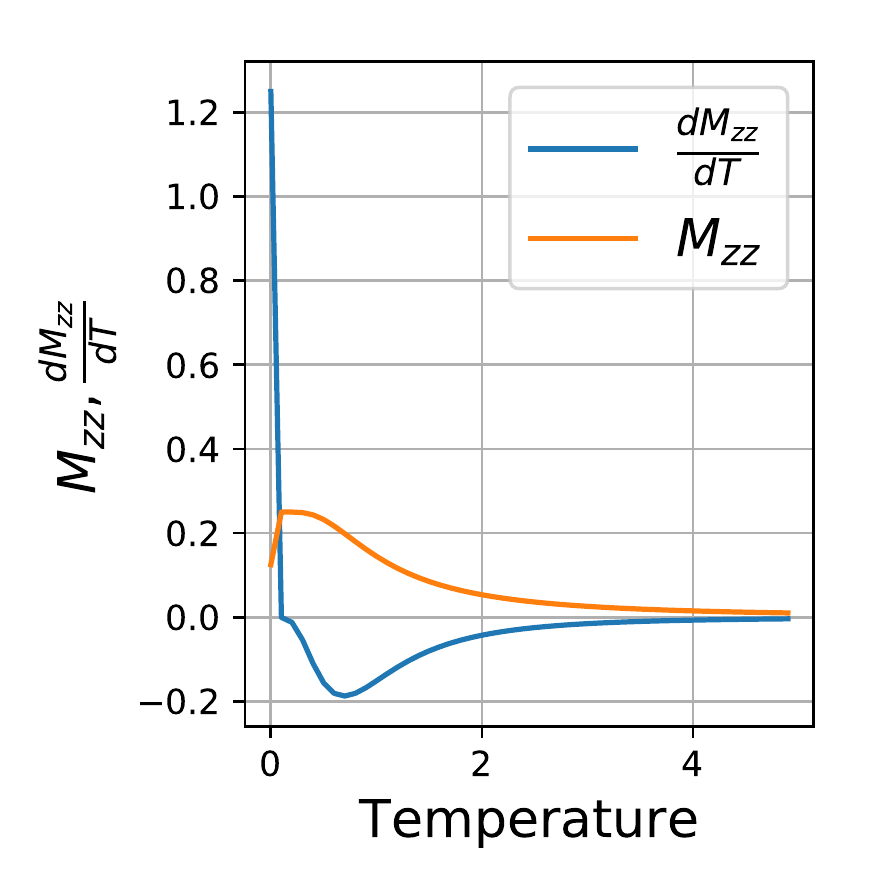}
\raisebox{3cm}{\large\textbf{(f)}}\hspace*{-1mm}
\includegraphics[width=0.28\linewidth]{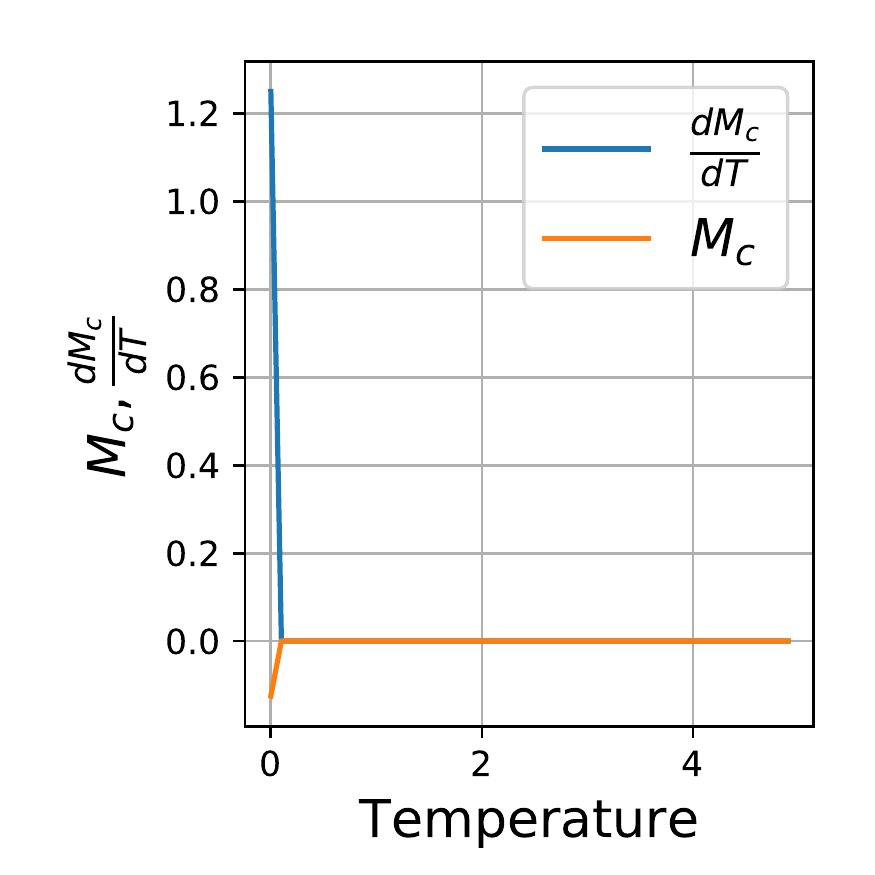}
\caption{(a), (b) and (c) show the plots of three observables $\{M_z, M_{zz}, M_c\}$ as a function of time
$(t)$ respectively, obtained by numerically solving Eq. (\ref{3eq}), for three
choice of $\{R_1,M_{\circ}, \alpha\}$. The parameter sets are chosen as 
$\{R_1 = 1.0, M_{\circ} = 1.0, \alpha=1.0\}$ which correspond to $T = 0$ or common bath, the other sets
$\{R_1 = 5.0, M_{\circ} = 0.8, \alpha=0.5\}$, and
$\{R_1 = 10.0, M_{\circ} = 0.8, \alpha=0.5\}$ correspond to
two non-zero temperatures $T_1<T_2$. The parameter sets are labeled by $T_{\circ}$ (for $T=0$), $T_1$ and $T_2$ with
colors blue, orange and green, respectively (color online).
(a) and (b) shows that the equilibrium values of $M_z$ and $M_{zz}$ are less at $T_2$ compared to that of at
$T_1$. For $T_2 < T_1$ the behavior is expected. We note that the $T = 0$ behavior shows an
anomaly, i.e., $M_z\vert_{T = 0} < M_z\vert_{T_{1}}$.
 (c) $M_c$ is non-zero for $\alpha=1$ (i.e., $T = 0$) which
clearly shows the persistent entanglement, whereas no such persistent entanglement survives at the finite
temperatures.  
(d), (e) and (f) show the steady
state value each observable and their temperature derivative versus temperature ($T$). The first
order derivatives of all observables diverge at the point $T \to 0$. } 
\label{fig-time-temp}
\end{figure*}

The correlation length of the bath is modeled as a monotonic function of the temperature, as mentioned
earlier. Since the phase transition happens at $T = 0$, we study the observables' behavior in Eq.
(\ref{3eq}) as a function of temperature. Figure \ref{fig-time-temp}(a-c) depicts the behavior of the
observables $\{M_z, M_{zz}, M_c\}$ as a function of the time for a set of fixed temperatures. At $T \to 0$ or
$\alpha = 1$, the expectation value of zero-quantum operator $M_c$ remains finite in the steady-state,
whereas it vanishes for $T>0$. The existence of $M_c \neq 0$ indicates the presence of the persistent
entanglement. The vanishing of the entanglement for $T>0$ is in line with the works of Huelga and others,
who showed that regular Markovian local environments lead to separable steady-states
\cite{huelga_non-markovianity-assisted_2012}. Figures \ref{fig-time-temp}(d-f) show the observables'
behavior and their derivatives as a function of the temperature. The temperature has been varied linearly,
and $\alpha$ was calculated using the relation between $\xi$ and $\beta$ mentioned earlier.  The sudden jump
of the observables and the discontinuity of their derivatives clearly show the first-order nature of this
DPT.

Hence, there exist a critical value of $T_{\circ}$
($T=0$), which is responsible for the symmetry breaking phase transition of the system.
\begin{eqnarray}
\lim_{T \to T_{\circ}}\, \frac{\delta}{\delta T}\text{Tr}[\rho_s^{ss}(T) (D)] =\infty
\end{eqnarray}  
The first order derivative w.r.t $T$ vanishes at $T \to T_{\circ}$. So, it is a first-order phase transition 
\cite{minganti_spectral_2018}.
\begin{figure}[htb]
\includegraphics[width=3.5in]{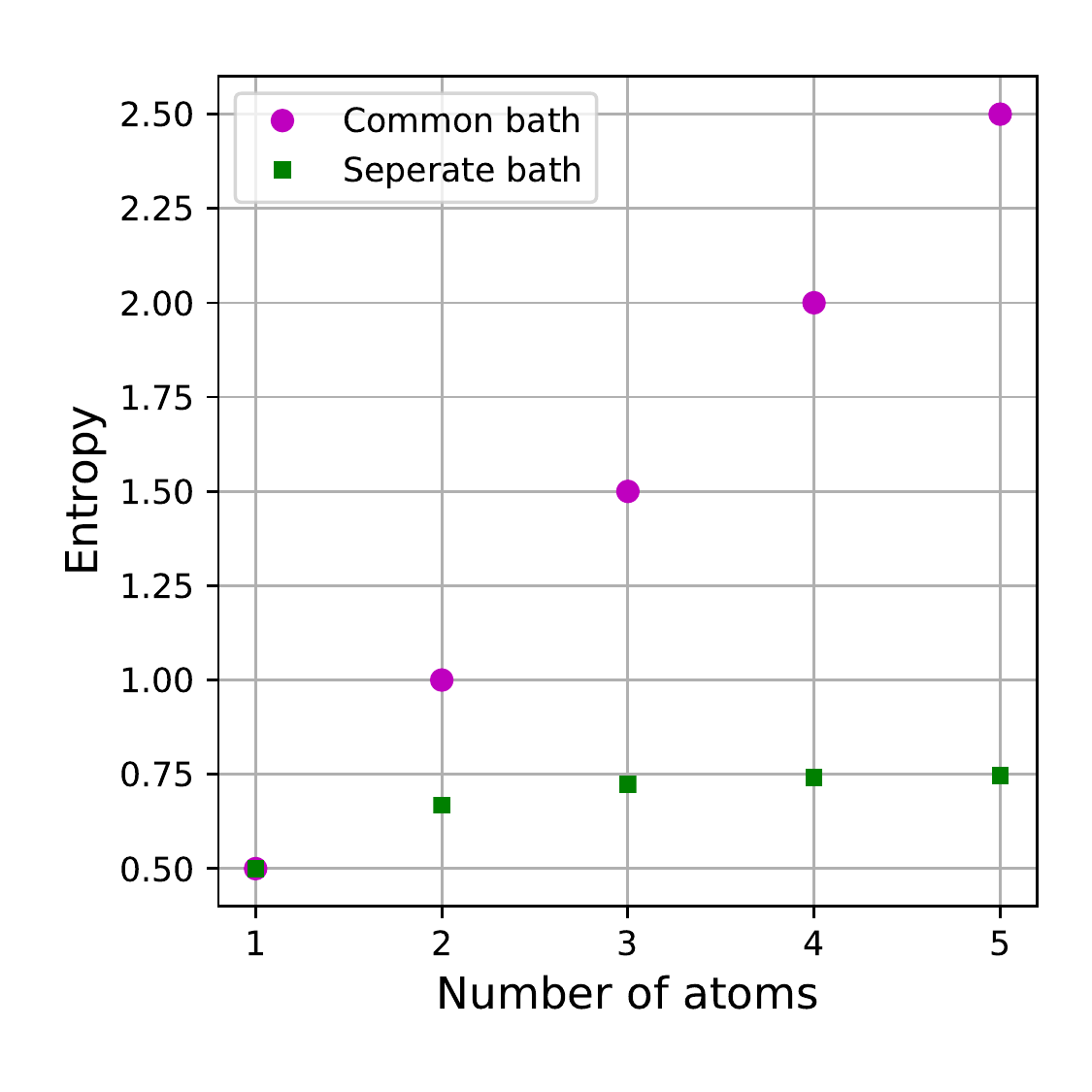} 
\caption{Figure shows the numerical plot of von Neumann entropy from the steady-state density matrix as a
function of the number of atoms.  To compare the effects of the common environment versus the separate
environment, we set $\alpha = 1$ (green filled-square markers for common environment) and $\alpha = 0.5$
(magenta filled-circle markers for
separate environment), while keeping other parameters, constant ($R_1=1.0$, and $M_{\circ}=0.6$).  It is
clear that for the separate environment ($\alpha < 1$), the entropy increases following the volume law and
hence is linear. On the other hand, for the common environment ($\alpha = 1$), the system is frozen and the
entropy is no longer an extensive thermodynamic quantity.}
\label{fig-entropy}
\end{figure} 

We note that since a common environment can exist at a finite temperature \cite{plenio_cavity-loss-induced_1999}, 
hence such phases do not strictly require one to reach zero temperature. Multiple ions have been confined in a
common electromagnetic field at a finite temperature using the trapped-ion technique
\cite{wineland_laser_2008, Poyatos1996}. Hence, for a longer bath-correlation length $(\xi)$, we can make
$\alpha \to 1$ at a sufficiently low temperature.  

We show here a comparative study of change of the von Neumann entropy
$(\mathcal{S}=-Tr_s\{\rho_s \ln\rho_s \})$ by increasing the number of atoms in a common bath and separate
bath, in figure \ref{fig-entropy}. The entropy is extensive for the separate local environments and
increases linearly with the number of constituent atoms ($n$).
But for common environments, one expects an area law obtained from the derivative of the volume law with
respect to $n$ and hence the entropy should be independent of $n$ \cite{eisert_colloquium_2010}.
We simulate the $n$ TLS, keeping them connected to a common environment. The switch over from
the volume law to the area law is shown in Fig. \ref{fig-entropy}.
Corresponding dark state, in terms of observables, can be calculated by putting $F=-3/4$,
we get $M_z=0$ and $M_c=-1/2$, $M_{zz}=-1/4$. The steady-state is temperature independent, and also it is a
pure state. The wave function of the dark state is, $\vert \psi \rangle= \frac{1}{\sqrt{2}}
(\vert 0   1 \rangle - \vert 1 0 \rangle )$, a singlet state between the two spins. With the addition of
other spins, such states are found for each pair of spins. An addition of another spin results in a set of
dark states. Upon further addition of spins, the entropy does not increase anymore (only the dark states are
created).

\section{Conclusion}

We identify a spatially-correlated bath as a completely common environment in the zero-temperature
limit. The same environment acts as a regular local environment with non-zero temperatures. Several
conserved quantities can be identified when non-interacting systems are in a common environment. Hence
persistent entanglement exists under this dissipative dynamics. Increasing the number of spins, the number
of integrability also increases. On the other hand, the presence of separately local environments ensures
that the system is non-integrable. Hence, the system evolves to a thermal state after a sufficiently long
time. There exist a temperature-driven first-order phase transition. At the critical limit ${T \to 0}$, a
weak symmetry characterizes the final steady-state, the system skips thermalization.  We note that recent
laser-cooled ion-trap experiments set a major goal to observe this effect at a finite temperature \cite{
poyatos_quantum_1996, wilson-rae_laser_2004, sarlette_stabilization_2011, tomadin_reservoir_2012,
stellmer_laser_2013}. 

\begin{acknowledgments}
The authors thank Arpan Chatterjee and Pragna Das for insightful discussions and
helpful suggestions. SS gratefully acknowledges University Grants Commission for a research fellowship
(No.: 1431/ (CSIR-UGC-Net DEC. 2017)).
\end{acknowledgments}
 
\bibliographystyle{apsrev4-1}
\bibliography{references}
\end{document}